\documentclass[10pt,a4paper,prl,reprint,superscriptaddress]{revtex4-1}

\usepackage[T1]{fontenc} \usepackage{pstricks} \usepackage{subfigure}
\usepackage[normalem]{ulem}
\usepackage{graphicx} 
\usepackage{amsbsy}
\usepackage{amsfonts}
\usepackage{amsmath}
\usepackage{amssymb}
\usepackage{fixmath}
\usepackage[colorlinks]{hyperref}
\usepackage{physics}
\usepackage{color}

\hypersetup{
	bookmarksnumbered,
	pdfstartview={FitH},
	citecolor={green},
	linkcolor={blue},
	urlcolor={darkblue},
	pdfpagemode={UseOutlines}}
\definecolor{darkgreen}{RGB}{20,100,20}
\definecolor{darkblue}{RGB}{0,0,130}
\definecolor{darkred}{rgb}{.8,0,0}

\newcommand{\nn}{\nonumber}
\providecommand*{\I}{\mathrm{i}} 

\renewcommand{\vec}[1]{\mathbold{#1}}
\renewcommand{\d}{\mathrm{d}}
\newcommand{\spinor}[3]{\left(\hspace{-1.0ex}\begin{array}{c}#1\\ #2\end{array}\hspace{-1.0ex}\right)_{\hspace{-0.7ex}#3}\hspace{-2.0ex}}
\newcommand{\nab}{\mathbold{\nabla}}

\begin{document}
\title{Unified description of dynamics of a repulsive two-component Fermi gas }

\author{Piotr T. Grochowski}
\email{piotr@cft.edu.pl}
\affiliation{Center for Theoretical Physics PAN, Al. Lotnik\'ow 32/46, 02-668 Warsaw, Poland}
\author{Tomasz Karpiuk}
\email{t.karpiuk@uwb.edu.pl}
\affiliation{Wydzia{\l} Fizyki, Uniwersytet w Bia{\l}ymstoku,  ul. K. Cio{\l}kowskiego 1L, 15-245 Bia{\l}ystok, Poland}
\author{Miros{\l}aw Brewczyk}
\email{m.brewczyk@uwb.edu.pl}
\affiliation{Wydzia{\l} Fizyki, Uniwersytet w Bia{\l}ymstoku,  ul. K. Cio{\l}kowskiego 1L, 15-245 Bia{\l}ystok, Poland}
\author{Kazimierz Rz\k{a}{\.z}ewski}
\email{kazik@cft.edu.pl}
\affiliation{Center for Theoretical Physics PAN, Al. Lotnik\'ow 32/46, 02-668 Warsaw, Poland}

\date{\today}

\begin{abstract}

We study a binary spin-mixture of a zero-temperature repulsively interacting $^6$Li atoms using both the atomic-orbital and the density functional approaches.
The gas is initially prepared in a configuration of two magnetic domains and we determine the frequency of the spin-dipole oscillations which are emerging after the repulsive barrier, initially separating the domains, is removed.
We find, in agreement with recent experiment (G. Valtolina et al., Nat. Phys. 13, 704 (2017)), the occurrence of a ferromagnetic instability in an atomic gas while the interaction strength between different spin states is increased, after which the system becomes ferromagnetic.
The ferromagnetic instability is preceded by the softening of the spin-dipole mode.

\end{abstract}

\maketitle

The interaction of itinerant fermions, i.e. the ones not localized in the lattice, causes a variety of effects in many quantum systems~\cite{Giorgini2008,Brando2016,Silverstein1969}.
Among them, the existence of ferromagnetism in binary spin-mixtures has been addressed numerous times in both theory~\cite{Sogo2002,Karpiuk2004,Duine2005,LeBlanc2009,Conduit2009,Cui2010,Pilati2010,Chang2011,Pekker2011,Massignan2011,Massignan2014,Trappe2016,Miyakawa2017} and experiment~\cite{DeMarco2002,Du2008,Jo2009,Sommer2011,Sanner2012,Lee2012,Valtolina2016}.
A simple mean-field treatment of the homogeneous electron gas, introduced by Stoner, predicts a ferromagnetic phase, when a short-ranged screened Coulomb repulsion between two particles of opposite spins dominates over the degeneracy pressure~\cite{Stoner1933}.
However, this idealized picture proved insufficient to satisfactorily describe majority of encountered phenomena as the effects beyond mean-field often play a crucial role~\cite{Brando2016}.
Still, clean fermionic systems in which short-range interaction is dominant are experimentally available with the help of Feshbach resonances.
Yet, the existence and stability of ferromagnetic phase in such systems are still in dispute.

The use of methods more sophisticated than Stoner approach likewise leads to ferromagnetic instability which can be driven by short-range interactions only.
Such a ferromagnetic state is however a metastable one, as it corresponds to the excited energy branch~\cite{Chin2010}.
It is due to the fact that the short-range repulsive interaction, ultimately described in terms of the s-wave scattering channel, calls for underlying attractive potential with a weakly bound molecular state~\cite{Chin2010}.

Indeed, recent experiments show that binary spin-mixture of Fermi gas, initially prepared in a paramagnetic state, decays to the superfluid state of paired fermions rather than creating a ferromagnetic phase~\cite{Sanner2012,Lee2012}.
To decrease the pairing rate, in the experiment~\cite{Valtolina2016}, $^6$Li atoms were prepared in an artificial ferromagnetic state, in which both components are initially separated by the optical barrier in a harmonic trap and then undergo a free time evolution after the release of the barrier.
The usual treatment of dynamics of such a spin-polarized system distinguishes between the weak~\cite{Duine2010,Pilati2010,Recati2011,He2012} and the strong interaction~\cite{Taylor2011,Palestini2012,Mink2012} regimes.
The only available unified descriptions are based on semi-classical Boltzmann equation~\cite{Toschi2003,Goulko2011}.

To tackle this problem in an alternative way, we employ two fundamentally different methods, namely atomic-orbital (time-dependent Hartree-Fock) and density-functional approaches.
Both of them provide the results for a very broad range of interaction strengths and quantitatively agree with each other and with the experiments~\cite{Sommer2011,Valtolina2016}.
We observe three regimes of the gas behavior with the increase of the interaction -- miscible, intermediate and immiscible.
For a weak coupling two fermionic clouds pass through each other and for the strong one the clouds bounce off each other.
In both these cases a single frequency of relative center-of-mass oscillations can be found.
However, this is not the case in the intermediate regime in which spontaneous creation of unstable domain structure takes place.

In the simple atomic-orbital description of a Fermi gas we apply, it is assumed that the many-body wave function $\Psi({\bf x}_1^{},...,{\bf x}_{N})$ of $N$ indistinguishable fermionic atoms is given by the single Slater determinant 
\begin{eqnarray}
&&\Psi ({\bf x}_1,...,{\bf x}_{N}) 
= \frac{1}{\sqrt{N!}} \left |
\begin{array}{lllll}
\varphi_1({\bf x}_1) & . & . & . & \varphi_1({\bf x}_{N}) \\
\phantom{aa}. &  &  &  & \phantom{aa}. \\
\phantom{aa}. &  &  &  & \phantom{aa}. \\
\phantom{aa}. &  &  &  & \phantom{aa}. \\
\varphi_{N}({\bf x}_1) & . & . & . & \varphi_{N}({\bf x}_{N})
\end{array}
\right |  .   \nonumber  \\
\label{Slater}
\end{eqnarray}
Here, the coordinates ${\bf x}_i$ of an atom comprise both spatial and spin variables and $\varphi_i({\bf x}),\, {i=1,...,N}$ denote different, orthonormal spin-orbitals.
We further assume that the spin-dependent part of spin-orbitals is twofold and that exactly half of the atoms occupy each spin state. 

At low temperatures atoms occupying each spin state can be considered as noninteracting Fermi gas.
The only interaction present in the system is the repulsion between atoms of different spins which we describe by a contact potential characterized by the coupling constant ${g\ge0}$, related to the $s$-wave scattering length $a$ through ${g=4\pi\,a\hbar^2/m}$.
Hence, the time-dependent Hartree-Fock equations for the spatial parts of the spin-orbitals, i.e. spatial orbitals representing the first component, $\varphi_{i,+} ({\bf r},t)$, and the second component, $\varphi_{i,-} ({\bf r},t)$, are given by
\begin{eqnarray}
&& i\hbar \partial_t  \varphi_{i,\pm} ({\bf r},t) =
( -\frac{\hbar^2}{2 m} \nabla^2 + V_{tr}({\bf r})  
+ g n_{\mp} ({\bf r},t) \; ) \; \varphi_{i,\pm} ({\bf r},t) \nonumber \\
\label{HFeq}
\end{eqnarray}
for $i=1,...,N/2$ and with
\begin{eqnarray}
n_{\pm} ({\bf r},t) = \sum_{j=1}^{N/2} |\varphi_{j,\pm} ({\bf r},t)|^2   \,.
\end{eqnarray}   
Note that these evolution equations preserve orthogonality of the orbitals.

We consider an equally populated binary spin-mixture of ultracold $^6$Li atoms.
We start our numerics assuming small numbers of atoms in each component, equal to $N/2=24$.
Just like in the experiment~\cite{Valtolina2016} the atoms are confined in the axially symmetric harmonic trap with axial and radial frequencies equal to $\omega_z=2\pi\times 21\,$Hz and $\omega_r=2\pi \times 265\,$Hz, respectively.
The system is initially prepared in the configuration of two fully separated domains.
It is done numerically by raising additional high enough barrier at the plane $z=0$ to prevent mutual interactions and determining $N$ lowest energy spatial orbitals for the ideal gas in such a modified trapping potential. 

\begin{figure*}[h!tbp]
	\centering
	\includegraphics[width=0.99\linewidth]{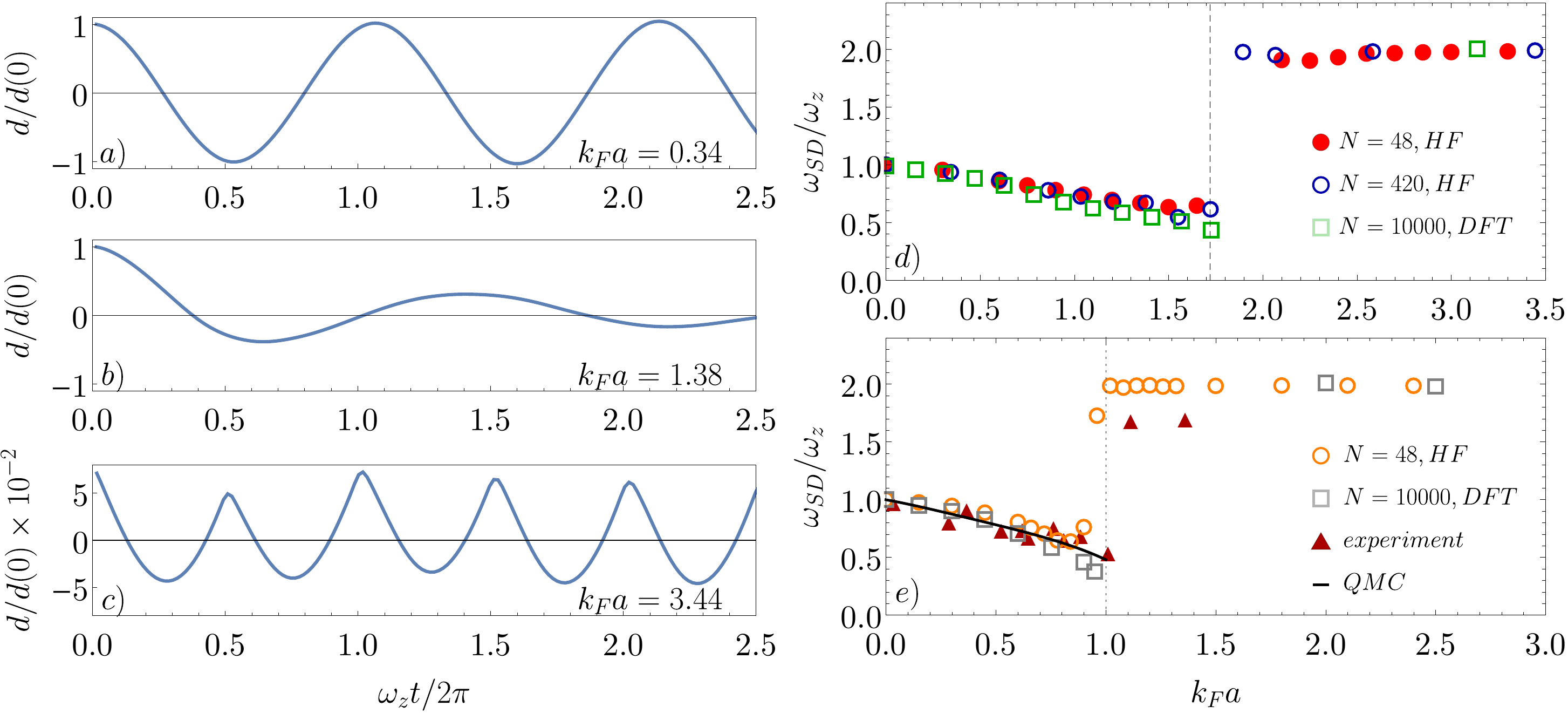}
	\caption{(a)-(c) Relative distance between the centers of mass of two atomic clouds $d(t)$ (an average value is subtracted) for the system of $N=420$ atoms as a function of time, calculated within time-dependent Hartree-Fock approach.
	With an increase of the interaction strength, the frequency of the spin-dipole mode gets lower (the softening of the spin-dipole mode) and the amplitude of the oscillations becomes damped.
	After a very narrow intermediate regime, for a strong interaction, two clouds become immiscible and the periodic behavior is revived as fast small-amplitude oscillations.
	(d)-(e) Frequency of the spin-dipole mode of a repulsive Fermi gas as a function of the interaction strength, calculated without (d) and with (e) renormalization of the scattering length.
	The ferromagnetic instability occurs at approximately $k_F a\approx 1.0$ if higher-order terms with respect to correlations are included in Eqs. (\ref{HFeq}) and (\ref{pseudopsi}).
	The solid line in (e) is the $T=0$ quantum Monte Carlo prediction from Ref.~\cite{Recati2011} assuming full overlap of both species.
	The red triangles are $T/T_F=0.12$ experimental data from Ref.~\cite{Valtolina2016} (presented in a slightly different variable, for its exact definition see Methods in Ref.~\cite{Valtolina2016}).
	Far beyond the critical point, the initial ferromagnetic phase becomes stable.
	The results for different sizes of the system are presented, revealing the universal behavior of the gas with respect to dimensionless variable $k_Fa$, calculated within different approaches.
	The difference in the size of the intermediate regimes is addressed in the text.\label{BigN}}
\end{figure*}

Hence, we start to study the dynamics of the gas which is initially in the ground state of the harmonic trap plus the separating barrier potential.
We abruptly remove the separating barrier and observe emerging oscillations of the atomic clouds.
Fig.~\ref{BigN} summarizes our results.
Frames (a), (b), and (c) show the time dependence of the relative distance between the positions of the centers of mass of two atomic clouds.
For a very small repulsion the atomic clouds almost do not see each other, as they oscillate with frequencies equal to the axial trap frequency.
When the repulsive interaction increases we observe the decrease of the frequency of the spin-dipole mode down to the value about $0.5\, \omega_z$ (see Figs.~\ref{BigN}(d,e)).
In this range of interactions the oscillations are strongly damped.
The response of the system changes qualitatively when the repulsion is further increased.
The atomic clouds become immiscible and oscillate with the frequency close to twice the axial trap frequency. 

Figs. \ref{BigN}(d,e) show the frequency of the spin-dipole mode in the wide range of repulsive interaction strengths.
The interaction strength is given in the dimensionless variable $k_F\, a$, where $k_F$ is the Fermi wave number ${k_F=(24 N)^{1/6}/a_{HO}}$, and $a_{HO}$ is the geometrical average of harmonic oscillator lengths (see e.g.~\cite{Recati2011}).
Evidently, there is a narrow region of interactions around $k_F\, a \approx 1.7$ (in fact, around $1.0$ -- see subsequent discussion) which separates two qualitatively different regimes.
For a weak enough repulsion two atomic clouds are miscible whereas for strong repulsion components get separated, i.e. the system enters the ferromagnetic phase.
Hence, the model of a two-component Fermi gas we use, features the ferromagnetic instability.
The critical value of $k_F\, a$ at which this instability occurs does not depend on the number of atoms in the system (see Fig. \ref{BigN}(d)) revealing its universal behavior as it should be according to Stoner's model of itinerant ferromagnetism~\cite{Stoner1933}.
Fig.~\ref{BigN}(d) shows the results for systems with the number of atoms equal to $N=48$ and $N=420$.
However, further increase of the number of particles proves to be very challenging numerically within the atomic-orbital approach.

In order to analyze systems closer to the experimental setups, we employ density functional approach introduced for such systems in Ref.~\cite{Trappe2016}.
This time, we consider a spherically symmetric trap to simplify the numerics and to provide a further check of the universality of the gas behavior under a different geometry.
Underlying energy functional that we consider goes beyond the usual Thomas-Fermi approximation as it includes gradient (Weizs\"{a}cker) corrections to the kinetic energy functional~\cite{Weizsacker1935,Kirzhnits1957}.
These corrections appear to be crucial for a construction of a reliable ground state.
The contact interaction term is given as an overlap between the density profiles of the components, $E_{\text{int}}=g \int  \d {\vec{r}} \, n_+ \, n_-$.
Such a treatment implies neglecting intercomponent correlations and assumes high occupation of both spin components.
The coupling constant ${g}$ is therefore a free parameter.
The time evolution of the system is then handled with the hydrodynamical approach~\cite{Madelung1927}.
Let us introduce the pseudo-wave function
\begin{align}\label{TwoCompPsi}
\psi=\spinor{\psi_+}{\psi_-}{\phantom{\Delta\tau}}=\spinor{\sqrt{n_+}\,e^{\I\frac{m}{\hbar}\chi_+}}{\sqrt{n_-}\,e^{\I\frac{m}{\hbar}\chi_-}}{\phantom{\Delta\tau}},
\end{align}
where ${n_++n_-=\psi^\dagger\psi}$ is the total one-particle density, and ${\nab\chi_\pm=\vec v_\pm}$ are the velocity fields of the collective motion.
The full system Hamiltonian is given by $H=T_{\text{tot}}+E_{\text{int}}+E_{\text{pot}}$.
The total kinetic energy ${T_{\text{tot}}=T+T_{\text{c}}}$ consists of the intrinsic kinetic energy $T$, which we approximate by the Thomas-Fermi-Weizs\"{a}cker functional, and the kinetic energy of the collective motion, ${T_{\text{c}}=\sum_{j=\pm}\int\d\vec{r}\  \frac{m}{2}n_j\,\vec v_j^2}$.
The potential energy is of the form $E_{\text{pot}}=\sum_{j=\pm}\int\d\vec{r} \  V_j n_j$.
Throughout this work we assume zero-curl velocity fields ${\vec v_\pm=\nab\chi_\pm}$.
By means of the inverse Madelung transformation, we obtain a nonlinear pseudo-Schr\"odinger equation, governing the time evolution of the pseudo-wave function:
\begin{align}\label{pseudopsi}
\I\hbar\partial_t\psi_\pm&= \Big[-\frac{\hbar^2}{2m}\nab^2+\frac{4\hbar^2}{9m}\frac{\nab^2|\psi_\pm|}{|\psi_\pm|}\nn\\
&\quad+A\,|\psi_\pm|^{4/3}+V_{tr}+g |\psi_\mp|^2\Big]\psi_\pm,
\end{align}
where $A=6^{5/3} \hbar^2 \pi^{4/3}/12m$.

Equations of this type can be readily solved by the split-operator methods~\cite{Gawryluk2017,Taha1984}, yielding real or imaginary time propagation of a given pseudo-wave function.
To recreate the initial state we follow the approach from the previous section, employing a half-trap potential as a basis for the evolution in imaginary time.
We arrive at two separated clouds with no significant overlap, successfully reproducing the results from the orbital approach.
As for the evolution in real time, special care is needed.
The term proportional to $\nab^2|\psi_\pm|/|\psi_\pm|$ in~(\ref{pseudopsi}) is not bounded from below as $|\psi_\pm| \rightarrow 0$ and can become significant due to the numerical noise even in regions where the density profile has a vanishing tail.
Therefore, we introduce an appropriate cut-off for this term -- it is equaled to $0$ when $|\psi_\pm| < 10^{-4}$.
To further stabilize the evolution, we add a velocity-dependent dissipation-like term to the split-operator method, which simplifies into a slight retardation of the effective potential (the value of this retardation is $3 \cdot 10^{-4}$ in harmonic oscillator units).

\begin{figure}[htb] 
	\includegraphics[width=0.99\linewidth]{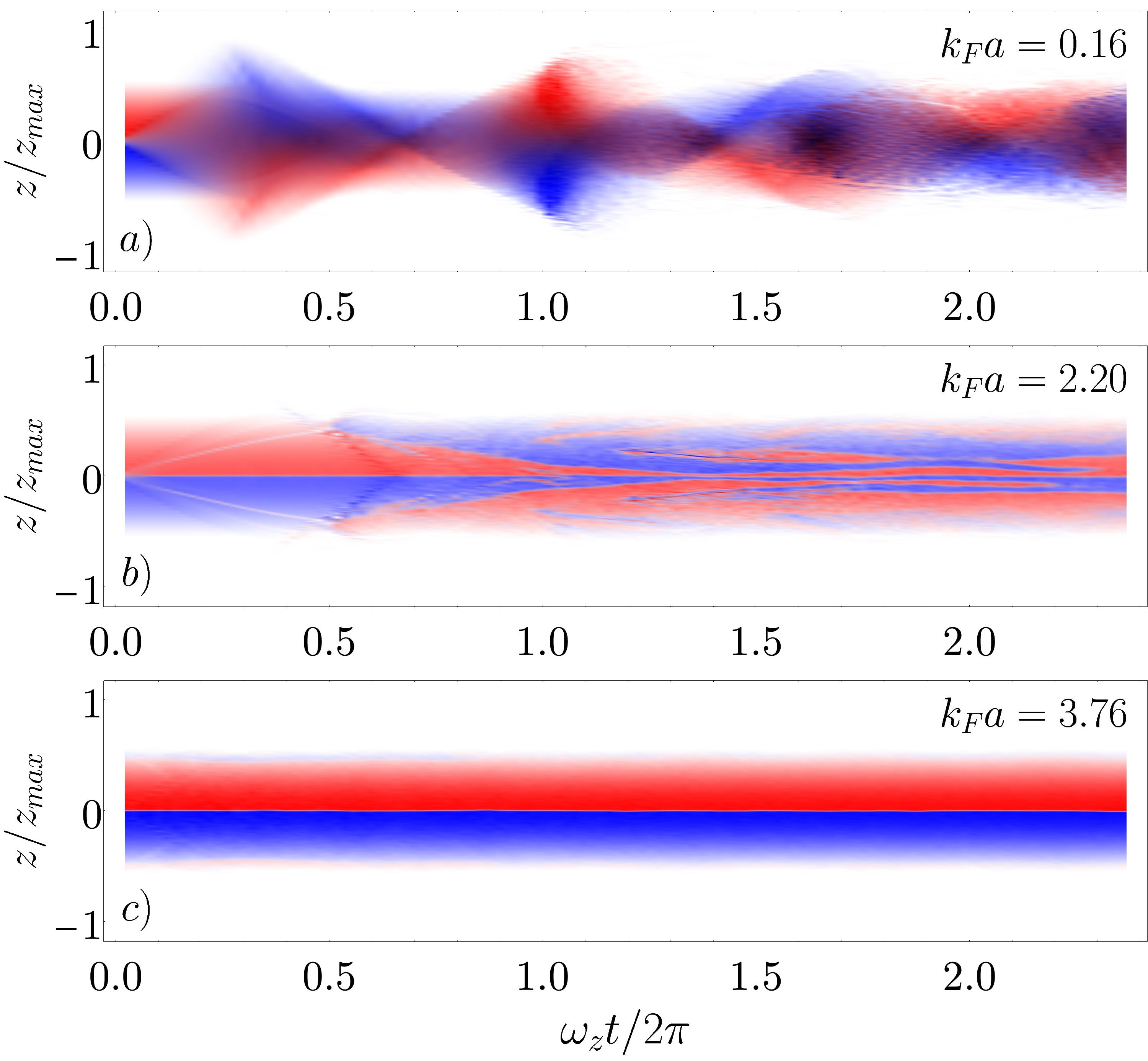} 
	\caption{(a)-(c) Density profile along the symmetry axis for each spin component as a function of time for different interaction strengths.
	For a weakly interacting gas (a), two clouds pass through each other in contrast to the strong interaction case (c), when two clouds are immiscible.
	The unstable domain structure in the intermediate regime (b) is clearly visible.} 
	\label{domains}
\end{figure}

We have performed the calculations for $N=1000$ and $N=10000$ particles, and obtained results that stay in agreement with the outcome of the previous method for smaller populations.
Once again, we observe three regimes (see Figs.~\ref{BigN}(d,e) and Fig.~\ref{domains}).
Softening of the spin-dipole mode occurs at weak interaction, going as low as ca. $0.5\, \omega_z$ near $k_Fa \approx 1.7$, before entering the regime in which no oscillatory motion of the clouds is visible.
For a strong interaction, initial domain structure is preserved and small-amplitude oscillations are observed, signaling the entry into the immiscible regime.

The size of the intermediate regime differs for HF and DFT approaches in Figs.~\ref{BigN}(d,e).
We attribute this discrepancy to the different trap geometries.
Contrary to the spherically symmetric case, in an elongated trap the gas transmission through the intercomponent interface on the perimeter is inhibited.
On the perimeter the gas density is smaller and as a result -- so is an overlap between two spin-components.
It means smaller local interaction energy and thus possibility of components passing through each other when it does not happen at the center of the trap.
The full transition into the immiscible regime is therefore shifted to the stronger interaction as a significant part of the gas is still allowed to move freely.

The universal behavior of a repulsive Fermi gas as encountered in our results can be understood through the Stoner instability~\cite{Stoner1933}.
In the simplified description, within the Thomas-Fermi approximation, the critical value of repulsive interactions can be found just by comparing the kinetic energy to the interaction one~\cite{Zwerger2009}.
For a uniform system it leads to $k_F\, a = \pi /2$.
This kind of instability yields different number of atoms in components.
Since in our model the number of atoms in each component is kept constant, the lowering of the interaction energy occurs rather via the separation of the atomic clouds (see Ref.~\cite{Trappe2016}).

There are two sets of data presented in Fig. \ref{BigN}.
Results obtained by solving Eqs. (\ref{HFeq}) and Eqs. (\ref{pseudopsi}) (and showing the critical value of $k_F a$ around $1.7$) only qualitatively agree with those reported in experiment \cite{Valtolina2016}, from which by extrapolation to zero temperature the critical value of $k_F a \approx 1$ is expected.
As predicted by the quantum Monte Carlo calculations \cite{Conduit2009,Pilati2010}, at zero temperature the ferromagnetic transition occurs at the interactions $k_F^b a \sim 0.8-0.9$ in the bulk, which corresponds to $k_F a \approx 1$ in the trap~\cite{Recati2011}.
A recent theoretical approach based on the dimensional epsilon-expansion method \cite{He2016} also leads to the similar critical value of $k_F^b a \simeq 0.79$.
The evident discrepancy between our model and the experimental results manifests because our description of the system is very simplified.
To improve the accuracy we must include in our approach the many-body correlations due to the interactions.
This can be done along the way prescribed in Ref.~\cite{Stecher2007}.
We renormalize the coupling parameter in the contact interactions locally, by replacing the bare scattering length, $a$, by the effective (and symmetrized) one: $a_{eff} = [\zeta{(k_+ a)}/{k_+} + \zeta{(k_- a)}/{k_-} ] /2$, where $k_+({\bf r})$ and $k_-({\bf r})$ are the local Fermi momenta for the first and the second component, respectively, and $\zeta{(\tilde{k}_F a)}$ is the renormalization function \cite{Stecher2007}.
We next expand the renormalization function $\zeta{(\tilde{k}_F a)}$ into powers of $\tilde{k}_F a$: $\zeta{(\tilde{k}_F a)} = \tilde{k}_F a + B (\tilde{k}_F a)^2 + D (\tilde{k}_F a)^3 + ... $.
The first term of the expansion is the usual scattering length for free two-particle scattering.
The second one with $B=6 (11 - 2 \ln 2)/35\pi$ was first obtained by Huang and Yang~\cite{Huang1957}.
It corresponds to the modification of the intermediate states by the Pauli exclusion principle.
The third term in general depends on specifics of the interatomic potential and includes three-particle correlations.
For hard sphere potential $D=1.084$~\cite{DeDominicis1957,Efimov1965}.
Keeping only first-order terms in the above expansion results in Eqs. (\ref{HFeq}) and (\ref{pseudopsi}).
However, taking into account the second- and third-order terms changes the time-dependent Hartree-Fock Eqs. (\ref{HFeq}) and pseudo-Schr\"odinger Eqs. (\ref{pseudopsi}) in the following way: $g n_{\pm} \rightarrow g n_{\pm} + C (4/3\, n_{\mp}^{1/3}\, n_{\pm} + n_{\pm}^{4/3} ) + E (5/3\, n_{\mp}^{2/3}\, n_{\pm} + n_{\pm}^{5/3} )$, where $C=ga B\, (6\pi^2)^{1/3}/2$, $E=g a^2 D\, (6\pi^2)^{2/3}/2$.
As seen in Fig. \ref{BigN}(e) the critical value of $k_F a$ is now around $1.0$ just as in theoretical considerations \cite{Conduit2009,Pilati2010,He2016} and as implied by the experiment \cite{Valtolina2016}.

The rate of small-amplitude oscillations within the immiscible regime measured in the experiment~\cite{Valtolina2016} is lower than our $T=0$ prediction.
This is likely due to temperature effects as the experimental data presented in Fig. 2(d) of Ref.~\cite{Valtolina2016} strongly suggest that for higher temperature this rate decreases.

In summary, we have presented two distinctly different descriptions of dynamics of the repulsive two-component Fermi gas in a wide range of interaction strengths.
Both of them quantitatively agree with the results of a recent experiment~\cite{Valtolina2016}.
In the limit of the ideal gas, both atomic clouds oscillate with the frequency equal to the trap frequency $\omega_z$.
For a weak repulsion we observe the softening of the spin-dipole mode, i.e. the difference of the centers of mass of the components starts oscillating with frequency smaller than $\omega_z$.
This softening effect is stopped at some critical repulsion.
This critical point occurs to be universal if expressed in terms of dimensionless parameter $k_F a$, staying independent of the number of atoms and of the trap geometry.
Hence, our calculations reveal the existence of the ferromagnetic instability in the two-component Fermi gas in which the pairing mechanisms are inhibited.
Beyond critical repulsion, for a strong interaction, the Fermi gas preserves its original domain structure and enters a stable ferromagnetic phase.

\acknowledgments  
P.T.G and K.R. were supported by (Polish) National Science Center Grant 2015/19/B/ST2/02820.
Part of the results were obtained using computers of the Computer Center of University of Bia\l{}ystok.

\end{document}